\documentclass[12pt]{article}  % authoryear,year
\usepackage{amssymb,graphicx} %,citesort} 
\usepackage {apalike}
\usepackage{mathptmx}% http://ctan.org/pkg/mathptmx  supposed to give the closest to Times New Roman

\begin{document}

%\begin{frontmatter}


%\date{21 November 2016} last revised

\date{Accepted 20 January 2017 \\ 
    \emph{Erkenntnis} {\bf 83} (2018)  pp. 135-161; doi:10.1007/s10670-017-9883-5.} % 

 \author{J. Brian Pitts \\ Faculty of Philosophy \\ University of Cambridge
\\ jbp25@cam.ac.uk  }  

\title{  Kant, Schlick and Friedman on Space, Time and Gravity in Light of Three Lessons from Particle Physics }   %Kant's Synthetic \emph{A Priori} Knowledge Rescued by Particle Physics from Schlick's General Relativistic Critique


\maketitle

\begin{center}
% \author{Running head:  Kant, Schlick and Friedman in Light of Particle Physics  }
\end{center}

%\pagebreak

\vspace{.2in} 

%%%%%%%%%%%%%%%%%%%

\begin{abstract}

 Kantian philosophy of space, time and gravity is significantly affected in three ways by particle physics.  First, particle physics deflects Schlick's General Relativity-based critique of synthetic \emph{a priori} knowledge. Schlick argued that since geometry was not synthetic \emph{a priori}, nothing was---a key step toward  logical empiricism.  Particle physics suggests a Kant-friendlier theory of space-time and gravity presumably approximating General Relativity arbitrarily well, massive spin-2 gravity, while retaining a flat space-time geometry that is \emph{indirectly} observable   at large distances. The theory's roots include Seeliger and Neumann in the 1890s and Einstein in 1917 as well as 1920s-30s physics.  Such theories have seen renewed scientific attention since 2000 and especially since 2010 due to breakthroughs addressing early 1970s technical difficulties.  

Second, particle physics casts additional doubt on Friedman's constitutive \emph{a priori} role for the principle of equivalence. Massive spin-2 gravity presumably should have nearly the same empirical content as General Relativity while differing radically on foundational issues.   Empirical content even in General Relativity resides in partial differential equations, not in an additional principle identifying gravity and inertia.    

Third, Kant's apparent claim that Newton's results could be known \emph{a priori} is undermined by an alternate gravitational equation. The modified theory has a smaller (Galilean) symmetry group than does Newton's.  What Kant wanted from Newton's gravity is impossible due its large symmetry group, but is closer to achievable given the alternative theory.

\end{abstract}  

%keywords:  scalar gravity, Klein-Gordon equation, Nordstr\"{o}m, massive, graviton, cosmological constant, conventionality, neo-Kantianism, Vienna Circle, Moritz Schlick, Henri Poincar\'{e} %, Hans Reichenbach, Arthur Eddington 

% \linenumbers

\pagebreak

%%%%%%%%%%%%%%%%%%%%%%%%%%%%%%%%%%%%%%%%%%%%%%%%%%%%%%%%%%%%%%%%%%%%%%%%%%%%%%%%%%%%%%%%%%%%%%%%%%%%%%%%%%%%%%%%%%%%%%%%%%%%%%%%%%%%%%%%%%%%%%%%%%%%%%

\section{Introduction}

Reliable scientific knowledge should not depend strongly on accidents, or at least on not accidents that lead us to misjudge how evidence supports our theories.  Failure of imagination can lead to our not entertaining theories that are comparably good to the ones that we did entertain; such unconceived alternatives undermine scientific realism
 \cite[p. 143]{vanFraassenLS} \cite{SklarUnborn,StanfordUnconceived,Roush,WrayUnderconsideration,KhalifaUnderconsideration,Underconsideration}. 
  This problem is rendered systematic by the fact that, as shown in Bayesianism, scientific theory testing is comparative \cite{ShimonyInference,EarmanBayes,SoberBook,IrrelevantConjunctionSynthese}. How well the evidence $E$ fits my favorite theory $T$ depends, perhaps surprisingly, on how likely some other theory $T_1$ (indeed all other theories, including ones that I haven't thought of) makes the evidence $E$ and how I spread my degrees of belief among the other theories $T_1,$ $T_2,$ \emph{etc.}  One sees this fact by expanding the denominator $P(E)$ of Bayes's theorem using the theorem of total probability:  $$P(E)= P(E|T) P(T) + P(E|T_1) P(T_1) + P(E|T_2) P(T_2) +\ldots.$$
 In the interest of freeing ourselves from historical accidents regarding space-time theory, it is prudent  to employ whatever systematic means exist for generating plausible alternative theories.  

Fortunately, there is a largely untapped source here, the literature that studies all possible classical (non-quantum) relativistic wave equations; it has gone untapped for a number of reasons, including a superficially quantum vocabulary.  That literature is particle physics, of which the late 1930s  taxonomy of relativistic wave equations in terms of mass and spin (\emph{e.g.}, \cite{PauliFierz,FierzPauli,WignerLorentz}) is  a prominent example.  In 1939 particle physicists Wolfgang Pauli and Markus Fierz  began to subsume Einstein's prematurely invented \cite[p. 334]{OhanianEinsteinMistakes}  General Relativity within the particle physics taxonomy  as massless and spin-$2$  \cite{PauliFierz,FierzPauli}.  Pauli and Fierz's  work also makes it natural to consider a small non-zero mass and spin-$2$ as a potential alternative theory, one which (as Seeliger understood in a simpler example already in the 1890s \cite{Seeliger1895a,NortonWoes}) presumably would approximate General Relativity as closely as desired.  This expectation was so overwhelmingly natural that its failure (at least with approximate calculations)  discovered in 1970  was a ``bombshell'' \cite{DeserMassSalam}.

%
% He wrote  (as translated by John Norton) that Newton's law was ``a purely empirical formula and assuming its exactness would be a new hypothesis supported by nothing.'' \cite{Seeliger1895a,NortonWoes}
%That claim is too strong, in that Newton's law had virtues that not every rival formula empirically viable in the 1890s had.  But a certain kind of exponentially decaying formula was associated with an appropriate differential equation  and hence had theoretical credentials comparable to Newton's \cite{PockelsHelmholtzEquation,Neumann}, vindicating the spirit of Seeliger's point.   
%The idea of exploring whether a massive theory could work in place of a massless one (or \emph{vice versa}), much as Seeliger proposed, is a commonplace in particle physics.

%%%%%%%%%%%%%%%%%%%%%%%%%%%%%%%%%%%%

\subsection{Particle Physics Background}

Pondering Maxwell's electromagnetism and Einstein's General Relativity, general relativists and philosophers often discuss relativistic wave equations in which the waves travel at the `speed of light.' 
In particle physics it is routine to consider also  wave equations for some particle/field(s), such as electrons, (some?) neutrinos, the weak nuclear force $W^{\pm}$ and $Z$ bosons, and maybe even light and/or  gravity  themselves,  that include  an \emph{algebraic} term in a field potential $\phi$ in the field equations.   The coefficient of such an algebraic term is the ``mass'' (squared) of the particle/field $\phi.$  Such terminology makes inessential  use of Planck's constant; I set the reduced version $\hbar$ to $1.$  The ``mass'' is in effect an inverse length scale, which one could take to be primitive, avoiding the appearance of Planck's constant (and the speed of light in that term). % PUT IN C AND HBAR
The resulting wave equation, invented multiple times around 1926 \cite{KraghKleinGordonManyFathers}, is generally known as the Klein-Gordon equation 
$$  \left(-\frac{1}{c^2} \partial^2/\partial t^2 + \nabla^2-m^2 c^2\right)\phi=0.$$ (Having displayed the speed of light $c$, I now set it to $1$ as well.)
``Particle mass'' in that sense is just a property of a classical field, an inverse length scale, expressed in entrenched quantum terminology for which there is no convenient alternative.
 In the static, spherically symmetric case, this equation becomes
$$ (\nabla^2-m^2)\phi= \frac{1}{r^2} \frac{\partial }{\partial r}\left(r^2 \frac{\partial \phi}{\partial r} \right) -m^2 \phi =0.$$ For a massive theory, one gets a faster, exponential fall-off as $\frac{1}{r}e^{-mr}$.  For wave solutions satisfying the Klein-Gordon equation, energy propagates (at the group velocity) more slowly than light at a speed(s) depending on $m$ and the frequency(s).

 A potential of the form $\frac{1}{r}e^{-mr}$ appeared in the 1890s in astronomy and physics independently in the work of Neumann and Seeliger \cite{Neumann1886,PockelsHelmholtzEquation,Neumann,Seeliger1896,Neumann1902Finite,NortonWoes,ScalarGravityPhil}  and again due to Yukawa in particle physics in the 1930s \cite{Yukawa}.  The inverse of $m$ is known as the range of the field, so nonzero $m$ gives a  field a finite-range, while $m=0$ gives a ``long'' or ``infinite'' range.   Seeliger and Neumann provided an alternative to Newton's theory by 1900, Seeliger providing cosmological motivations to make the gravitational potential converge in an infinite homogeneous universe and Neumann providing an appropriate partial differential equation and its solution.  Neither had  much to say about the physical meaning of the new parameter.  That lack of physical meaning and connection to other experience was noticed and faulted by Schlick \cite[p. 70]{Schlick}.  That lacuna was filled in the 1920s, however, making that aspect of Schlick's critique obsolete quickly, at a time when the contest between broadly Kantian and positivist conceptions of philosophy was still live.  That altered situation in physics unfortunately went unrecognized in philosophy, however.

In the late 1930s Pauli and Fierz found that the theory of a non-interacting massless spin 2 (symmetric tensor) field in Minkowski space-time was just the linear approximation of Einstein's General Relativity \cite{PauliFierz,FierzPauli}. 
Inspired by de Broglie and Pauli-Fierz, Marie-Antoinette Tonnelat and G\'{e}rard Petiau explored massive graviton theories on a sustained basis during the 1940s (cited below). % \cite{TonnelatFlux,PetiauCR41a}.  
Massive theories are plausible in terms of relativistic field theory.  Such work reached maturity (partly in other hands) in the 1960s \cite{OP,FMS}.
As Freund, Maheshwari and Schonberg put it, 
\begin{quote}
In the Newtonian limit, equation (1) is now replaced by the Neumann-Yukawa
equation, 
$$ (\Delta -m^2)V = \kappa \rho \hspace{1.2in}  (3),$$
which leads to the quantum-mechanically reasonable Yukawa potential 
$$ V(r) = - \frac{ \kappa M e^{-m r} }{r}, \hspace{1in} (4) $$\ldots
\cite{FMS}. 
\end{quote} 
This potential was sufficiently plausible as to be independently invented 3 times (Seeliger among many other potentials, Neumann, and Einstein in 1917 on the way to inventing his cosmological constant $\Lambda$ \cite{EinsteinCosmological}). Seeliger and Einstein were both addressing the problem of mathematically divergent gravitational potential in an infinite homogeneous static Newtonian universe.  Unfortunately philosophy never paid attention to massive spin-$2$ gravity, and hence failed to realize that Einstein's theory had serious competition (in the sense of decent prior probability, rendering the data nearly as likely as General Relativity did for all that anyone knew, and making a radical conceptual difference to space-time philosophy---a sort of philosophical expected utility) up to 1970 at least.  
Massive photon theories are fine even when merged with quantum mechanics to obtain massive quantum electrodynamics \cite{BelinfanteProca} (for many references see \cite{UnderdeterminationPhoton}).  

However, in the early 1970s, massive gravitons, which one would expect to behave analogously, ran into serious trouble on detailed technical grounds \cite{vDVmass1,vDVmass2,DeserMass}.  It was concluded that every such theory suffered from either an empirical problem due to a discontinuous massless limit $m \rightarrow 0$ (for pure spin $2$) or a problem of violent instability (for spin $2$ and spin $0$ together, because the spin $0$ (scalar) degrees of freedom have negative energy, so in quantum field theory one would expect explosive spontaneous creation of positive-energy spin $2$ and negative-energy spin $0$ gravitons).  Moreover, a theory that appeared to suffer from the former problem in the lowest approximation turned out to have the latter problem (in addition or instead) when treated exactly \cite{TyutinMass,DeserMass}, a problem recently dubbed the ``Boulware-Deser ghost.''   
Philosophers and historians who take the General Relativity side of the General Relativity \emph{vs.} particle physics rift in physics  \cite{Feynman,Rovelli}---which is most of them, often perhaps unwittingly---had gotten lucky.  The serious rival theories that they never contemplated, turned out not to work after all, the 1970s showed (or so it seemed).  Sometimes what you don't know won't hurt you.

It's not a reliable principle of scientific method, however, and in this instance  much of the original evidence has collapsed.  The tide has turned and massive graviton theories have been widely studied lately by physicists, who now know much about how to solve both the empirical discontinuity problem (partly with the help of numerical simulation)  \cite{Vainshtein,Vainshtein2} and how to solve the instability problem  \cite{deRhamGabadadze,HassanRosen,HassanRosenNonlinear,HinterbichlerRMP}. %\footnote{Amusingly, it turns out that part of the 2010 breakthrough solving the instability problem was already known in 1971, but no one cited that paper except its author, once, until now \cite{MaheshwariIdentity,MassiveGravity3}.} 
The competition between General Relativity (self-interacting universally coupled massless spin-$2$)  \emph{vs.} massive gravity (self-interacting universally coupled massive spin-$2$) is a well motivated example of the fact, noted by Pierre Duhem, that the curve fitting problem always applies in physics: through any set of experimental results, multiple curves can be proposed  \cite{Duhem}.

%%%%%%%%%%%%%%%%%%%%%%%%%%%%%%%%

\subsection{Outline} 

This paper will discuss three interrelated themes involving the surprising relevance of particle physics, in particular massive theories of gravity, to the well-studied Kant-Einstein interface.  First, Schlick's critique of Kant in light of General Relativity will be seen to be less than compelling once one clearly entertains the possibilities recognized in particle physics.  These possibilities partly predated Schlick's critique of Kant, but were fully developed later because Einstein developed his field equations earlier than the natural development of physics should have produced.  Second, Friedman's invocation of a constitutive \emph{a priori} role for the principle of equivalence, which has been criticized already by Don Howard \cite{HowardFriedman}, will be seen to be less than compelling in light of particle physics.  Particle physics shows how to construct a theory that, one would expect,  distinguishes inertia from gravitation, while empirically approximating Einstein's equations as closely as one wishes.   Third, though the symmetry group of Newtonian physics is much larger than Kant recognized, particle physics provides an alternate theory that reduces that gap.

%%%%%%%%%%%%%%%%%%%%%%%%%%%%%%%%%%%%%%%%%%%%%

\section{Massive Gravity \emph{vs.} Schlick's Critique of Kant from  General Relativity}

Moritz Schlick, future leader of the Vienna Circle, argued around 1920 that General Relativity made even a broadly Kantian philosophy of geometry impossible because the physical truth about the actual world was incompatible with it \cite{Schlick,SchlickCritical,Coffa,FriedmanKantKuhn,Ryckman}.
If even geometry is not an example of synthetic \emph{a priori} knowledge, then nothing is.  Ryckman has usefully framed the widely accepted view of the destructive significance of General Relativity for Kantian philosophy:  
\begin{quote}  
Kantian and neo-Kantian publications comprised a not-insignificant torrent in the
``relativity rumpus'' following the announced confirmation of the general theory of
relativity in November 1919. [footnote suppressed]  \ldots [I]t was incontrovertible that general relativity,
on corroboration of the dramatic prediction of star images displaced by the sun's
gravitational field, minimally required modification or clarification of the necessarily %p. 14 
 Euclidean structure of space implied by the Transcendental Aesthetic.  
Most of this literature, regardless of its provenance, contains little of present interest. But
within a few months in late 1920 and early 1921, Ernst Cassirer and Hans Reichenbach
published neo-Kantian appraisals of the theory of relativity whose historical and
philosophical significance has acquired renewed relevance at the beginning of the
21st century. [footnote suppressed]  \cite[pp. 13, 14]{Ryckman} \end{quote}
Some of these nameless Kantians and neo-Kantians are discussed by  Howard \cite{HowardEinsteinKant,HowardFriedman}. 
Cassirer's and Reichenbach's appraisals involve massive retrenchment.  Was that really necessary due to scientific progress?

There is a sense in which philosophers should not be expected solve that problem. It just isn't part of their professional training or responsibility to propose scientific theories, so if scientists don't propose the theories that philosophers need, who will?  One might hope that  % 
 someone knowledgeable about both subjects (perhaps someone like Schlick?) would take up that task.  But Schlick, as will appear, was too partisan to fill that role.  Someone who paid attention to what one could start to identify as particle physics in the 1920s (perhaps de Broglie or one of the many inventors of the Klein-Gordon equation \cite{KraghKleinGordonManyFathers})  could have given Kantian philosophers a friendly tip.  But that didn't happen, either, it seems \cite{Hentschel}.  Thus the available scientific resources for maintaining a Kantian position simply went unrecognized for a very, very long time, long past the time that many people cared about a Kantian position, in fact.

The question of the degree to which the progress of science is inevitable or contingent has received some attention \cite{HackingInevitable,SolerContingentScience2}.  What has not been noticed is that Ryckman's  widely shared assessment of the impact of General Relativity on the family of Kantian philosophies, in retrospect, was an historical accident.  Lakatos's point that  the actual contingent history must be held to normative standards in order to discern scientific progress \cite{LakatosFalsification,LakatosHistory} is borne out especially in the context of the problem of unconceived alternatives or underconsideration (\emph{e.g.}, \cite{StanfordUnconceived}). 
Indeed even a real historian of science, Kuhn (as opposed to Lakatos), agreed with Lakatos that historians should be prepared to identify historical actors' mistakes and that doing so was often important and illuminating \cite{LakatosKuhn}.

%%%%%%%%%%%%%%%%%%%%%%%%%%%%%%%%%%%

\subsection{Did Light Bending Verify General Relativity?} 

 It is striking how routinely the 1919 observed bending of light is construed as a \emph{verification} of General Relativity.  Impressive confirmation, yes, but verification?  That is a methodological holdover from 19th century Baconianism, contrary to Duhemian underdetermination, Popperian falsificationism, the promising parts of logical empiricist confirmation theory, Bayesianism, \emph{etc.} While one can forgive the enthusiasm of the popular media in 1919 in the aftermath of the Great War \cite{PaisEinstein}, the  \emph{New York Times} should not be allowed to  distort space-time philosophy permanently.  Yet  philosophers of science have been slow to apply standard philosophy of science ideas to the bending of light. It is fair to say that the bending of light  falsified Nordstr\"{o}m's scalar  theory of gravity \cite{Kraichnan,ScalarGravityPhil}.  Using 1920s-30s mathematics, one  can make the failure to bend light in Nordstr\"{o}m's theory \cite{EinsteinFokker,vonLaueNordstrom} manifest in that the part of the effective space-time geometry that light sees is untouched by gravity according to Nordstr\"{o}m's theory \cite{ScalarGravityPhil}. (It is difficult to imagine a plausible Duhem-Quine rescue story in this case.)

But surely other theories, perhaps not yet proposed and possibly not so revolutionary, might also predict the bending of light, so that it could not verify General Relativity? So, evidently, many physicists reasoned in the 1920s \cite{BrushLightBending}. It appears that for many people, including Bertrand Russell,  Whitehead's theory \cite{Whitehead,EddingtonWhitehead} filled that role \cite{RussellMatter}.  Massive gravity should have filled that role, especially from 1939 onwards. Whitehead's is not even a field theory, but rather a theory with retarded action at a distance in flat Minkowski space-time. Despite  the theoretical backwardness of action at a distance, Whitehead's theory was empirically viable still in the 1950s \cite{SchildWhitehead}.   It is wholly  appropriate that a role for a more conservative  theory of gravity and space-time than Einstein's was recognized, but filling it with only Whitehead's theory made the philosophical theses made to hang upon that role needlessly fragile.

Another source of confusion  is Einstein's mistaken analogy between his cosmological constant $\Lambda$ and his 1917 reinvention of the Seeliger-Neumann finite-range modification of Newtonian gravity  \cite{Heckmann,Trautman,Schucking,HarveySchucking,NortonWoes,EarmanLambda}. 
This faulty analogy has deceived many serious writers for a long time, including North  \cite[p. 179]{North}  \cite[pp. 515, 516]{NorthNorton}, Jammer  \cite[pp. 194, 195]{JammerSpace}, Pais \cite[p. 286]{PaisEinstein},  and Kragh  \cite[p. 28]{Kragh}.  
The mistake  happens in the middle of Einstein's cosmological constant paper, between the scalar and tensor sections.  
 That false analogy almost certainly helped to delay the conception of massive spin-$2$ gravity by decades. How could one think to do for the first time what Einstein supposedly had already done?   Or does Einstein's theory win forever because it was the \emph{first} tensor theory of gravity, the kind of theory that can bend light?

%%%%%%%%%%%%%%%%%%%%%%%%%%%%%%%

\subsection{Schlick's Contribution} 

Howard has outlined several useful themes:
\begin{quote}  
Schlick was one of the first philosophers to consider carefully the philosophical implications of relativity, and in Einstein's opinion, his analysis was far superior to those of most other philosophers because he did not try to appropriate the relativity theory to a partisan philosophical cause (as many neo-Kantians and positivists had done). Schlick also brought to his work a better grounding in physics than most other philosophers of his day could claim, for in 1904 he had taken a degree in physics under Max Planck at Berlin. Some years later, Schlick became the logical positivist we know as the founder of the Vienna Circle. But in 1915 Schlick's philosophy of science was a novel combination of realistic and conventionalistic components.

Schlick's first essay on relativity (1915) was published late in the same year in which Einstein completed his work on general relativity. The main purpose of the essay was to criticize the neo-Kantian and positivistic misinterpretations of relativity, and to exhibit, by way of contrast, some of the main philosophical implications that would be revealed by an unprejudiced
reading of the theory. \cite{HowardEinsteinSchlick}  \end{quote} 

I suggest that  Schlick felt less temptation to force-fit an alien philosophy onto General Relativity than did other philosophers because his philosophy already was more or less congenial to General Relativity. Einstein and Schlick had been drinking from many of the same wells, including Mach.  Perhaps Schlick's superior interpretation of General Relativity owes somewhat less to fair-mindedness than might otherwise be inferred, however.  To ascertain whether Schlick was less partisan as a philosophical commentator on space-time physics, one should look more broadly:  how well did he use his superior knowledge of physics to illuminate the philosophical discussion? 
More subtly, did he help other philosophers scientifically in a way that he was unusually, perhaps uniquely qualified to do? Did he notice philosophically interesting holes in the scientific literature and fill them---even if not congenial to his own philosophical projects?  Or did he use his expertise to claim beyond desert that science supported his philosophy?  Lawyers are unusually persuasive at making arguments, but in an adversarial system the prosecutor or defense attorney (unlike a judge) only takes one side.

As it turns out, Schlick's making plausible that General Relativity refutes Kant was an accident.  It depended crucially upon Schlick's  philosophically partisan failure to  apply his training as a physicist to ascertain whether the views that he wanted to undermine could be defended from his criticisms. Had he (or someone else) thought to propose it, it would have been easy to do to Einstein's theory what Seeliger, Neumann, and recently Einstein himself in 1917 \cite{EinsteinCosmological} had done to Newton's theory, thereby producing a Kant-friendlier theory that was presumably practically empirically equivalent to General Relativity.  Somewhat similar conceptual ingredients were available from Lotze \cite{Lotze}, who seems to have been curiously ignored in the relativity debate \cite{Hentschel} despite having said Poincar\'{e}-like things in defense of Kant on geometry well before Poincar\'{e}  \cite[288, 289, 408]{TorrettiPhilosophy} \cite[pp. 248, 249]{Lotze} \cite{Poincare}. Lotze, however, had no scientific theory and rendered the true geometry of space undetectable, as opposed to merely locally undetectable but indirectly globally discernible.    Analogous moves to Seeliger and Neumann soon would be made in electromagnetism by de Broglie, Proca and others in the 1920s-30s.  Such a proposal would not, of course, have suited Schlick's revolutionary project.  His anti-Kantian message and pro-Machian sympathies would be have been muted by even a hint of the possibility  of a modification of Einstein's equations which would approximate General Relativity as closely as desired, but containing a flat background metric that was observable in principle on astronomical scales and hence clearly real (albeit largely obscured by the distorting effects of gravity). Galileo often ignored Tycho; would Galileo have invented Tycho's theory out of fairness if Tycho hadn't already done so?  Why look for alternatives  that reduce the anti-Kantian sting of General Relativity to that of mere Special Relativity, when one has a partisan stance to defend?  There were progressive cultural implications in a Weimar climate of controversy and incipient reaction \cite{OkruhlikNeurathFeministValues}. 

  If neo-Kantian philosophy of geometry was overthrown (historically-sociologically at least) by General Relativity  due to arguments akin to Schlick's, then this is a good  example of the problem of unconceived alternatives.  The main problem for  Kantian philosophy was a \emph{lack of timely love from good physicists} who could have  proposed a scientifically serious and  Kant-friendlier theory of space-time and gravity. In the heyday of the debate there was not much chance that philosophers could identify the potential philosophical utility of incipient particle physics when even physicists had not done much in that direction. Much later but still long ago by now, things were different \cite{FMS}. 

 The untimely death of Poincar\'{e} (1912, age 58) is worth recalling here; it is easy to imagine Poincar\'{e} making such proposals in partial vindication of his conventionalist philosophy of geometry \cite{PoincareFoundations,WalterSchlickPoincare}.  It appears that much of  20th century philosophy was degraded by Poincar\'{e}'s death.  While people continued to talk about him, apparently no one followed him in a way that was both faithful and intelligent. Eddington, who had written a glowing obituary of Poincar\'{e} \cite{PoincareObituaryEddington}, later changed allegiance to Einstein and singled out only the most vulnerable parts of Poincar\'{e}'s view as representative \cite{EddingtonSTG}.  Logical empiricists could call themselves conventionalists but differed from Poincar\'{e} on a number of points, not always for the better.  Dingler's scientific intransigence \cite{TorrettiDingler} made his profession of conventionalism  more a liability than an asset.

It is widely believed that Schlick's work deploying General Relativity against neo-Kantian philosophy of geometry was both first-rate scientifically and philosophically at the time and of lasting significance.  Alberto Coffa  thought so: 
\begin{quote} Schlick was probably the first major philosopher to \emph{draw} the philosophical lessons of relativity.\ldots  Now the theory of [general] relativity had \emph{forced} his attention to the question of whether there is an apodictic \emph{a priori}.  A \emph{careful, prolonged} analysis  of the situation finally led him to conclude that there is no such thing and, more importantly, that this would \emph{entail} a decisive break with the Kantian tradition.  Schlick was the first one of the scientifically oriented neo-Kantians to \emph{understand} that the philosophical lessons of relativity demanded not the correction but the elimination of Kantianism.  \cite[pp. 196, 197]{Coffa} (emphasis added to highlight success terms) \end{quote} 
  This is the language of a monument to a lasting achievement.  

I suggest a different picture of this part of Schlick's work:  scientifically serviceable in its own time, but  partisan in using his scientific credibility to advance a philosophical agenda without making a scientific critique that he but not his opponents were capable of making---and also obsolete during the 1920s (not that this was pointed out).  Thus he created a facade that his philosophical opinions were entailed by scientific results, when instead he could have proposed a partly new theory using off-the-shelf ingredients that would have leveled the playing field.  His breathless endorsement of Einstein at the start of his own book \cite{Schlick} and hymn to the universe, General Relativity and Einstein (below) \cite[pp. 74, 75]{Schlick} do not suggest that the author was much interested in the epistemic caution involved in cultivating alternate theories.  He was far from Popper's critical spirit.  Furthermore, his work was obsolete when massive spin 2 gravity and the recognition of General Relativity as a massless spin $2$ theory  \cite{PauliFierz,FierzPauli} made it obvious how to start writing down a theory that, one would expect, would fit the data as well as Einstein's while having a fixed \emph{a priori} background geometry that is, in principle, observable astronomically. There were clues already in 1917 and more in the 1920s.

%%%%%%%%%%%%%%%%%%%%%%%%%%%%%%

\subsection{Lindemann's Challenge, Answered} 

One gets clear answers to the issues raised in the introduction by F. A. Lindemann for the English translation of Schlick, issues for philosophers, presumably especially Kantians, who wanted to preserve more traditional views about space and time. Here is Lindemann's challenge. 
\begin{quote} 
The main achievement of the \emph{general} theory of relativity has caused almost more difficulty to the school of philosophers, who would like to save absolute space and time, than the welding of space and time itself.  Briefly this may be stated as the recognition of the fact that it is impossible to distinguish between a universal force and a curvature of the space-time-manifold, and that it is more logical to say the space-time-manifold is non-Euclidean than to assert that it is Euclidean, but that all our measurements will prove that it is \emph{not}, on account of some hypothetical force.\ldots

At first sight it might appear that there must be an easy way to settle the question.  The golfer [who finds that balls  spiral into the hole, despite his inclination to believe the green level] has only to fix three points on his putting-green, join them by straight lines, and measure the sum of the three angles between these lines.  If the sum is two right angles the green is flat, if not, it is curved.  The difficulty, of course, is to define a straight line.  If we accept the definition of the shortest line, we have carried out the experiment, for the path of a ray of light is the shortest line and the experiment which determines its deflection may be read as showing that the three angles of the triangle---star---comparison star---telescope---are not equal to two right angles when the line star-telescope passes near the sun.  But some philosophers appear not to accept the shortest line as the straight line.  What definition they put in its place is not clear, and until they make it clear their position evidently is a weak one.  It is to be hoped they will endeavour to do this, and to explain the observed phenomena rather than adopt a merely negative attitude. \cite[pp. iv-vi]{Schlick} \end{quote}   
That was a well-framed and intellectually reasonable view in 1920, though not compelling for all rational beings. Whether it was \emph{professionally} reasonable to demand that philosophers make a novel contribution to physics and mathematics is harder to say, unless they were brought up as physicists first like Schlick.  Levi-Civita's general bimetric geometry, for which such nameless philosophers seem to have been groping, still lay in the future \cite{LeviCivita} (though many special cases involving flat and conformally flat geometries already were known \cite{ScalarGravityPhil}, and they suffice to make the conceptual point, though not to express an adequate theory of space-time and geometry in 1920).  But by now Lindemann's requests have been fulfilled, if not by philosophers, then by the particle physics tradition, and some of the key materials for doing so existed well before Lindemann wrote.  The massless spin $2$ derivations of Einstein's equations from flat space-time would eventually \cite{Kraichnan,Gupta,Feynman,Weinberg64c,Weinberg64d,Deser,SliBimGRG} show why it might not be unreasonable to favor universal forces even given Einstein's equations because it isn't implausible that gravity would act in just that way without any peculiar premises. 
(Such derivations turn out to be built around what one can recognize as the converse of Noether's Hilbertian assertion \cite{ConverseHilbertian}.) 
 Better yet, the already extant Neumann-Seeliger-Einstein 1890s/1917 modification of Newtonian gravity in principle showed the way to taking  massive spin-$2$ gravity (not a hypothetical force) to be an \emph{almost}-universal force   \cite{OP,FMS}, in the sense of acting like a Poincar\'{e}-Reichenbach universal force if one is unable to perform experiments sensitive to long-range gravitational effects.   Like their contemporary the Ford Model T, 
Lindemann's views, if construed as unanswerable rhetorical questions, are now somewhat dated.

%%%%%%%%%%%%%%%%%%%%%%%%%%%%%%%%%%%%%%%%%%%%%%%%%%%%%%%%%%%

\subsection{Neglect of Lotze}

Recall that General Relativity is viewed as making it 
\begin{quote} incontrovertible that general relativity,
on corroboration of the dramatic prediction of star images displaced by the sun's 
gravitational field, minimally required modification or clarification of the necessarily %p. 14 
 Euclidean structure of space implied by the Transcendental Aesthetic.   \cite[pp. 13, 14]{Ryckman}  \end{quote} 
If such scientific assistance as I have envisaged had been available, as it easily could have been, then not much clarification of Kant's philosophy would have been required beyond that already achieved in embryo by Lotze in his brief best moments  \cite[pp. 248, 249]{Lotze}, on top of whatever adjustments were required in updating Kant from space to space-time to fit special relativity.   According to Torretti, 
\begin{quote}
Lotze was, as far as I know, the first one to make the following important remark, which Poincar\'{e} later used in support of conventionalism.  In Euclidean geometry, the three internal angles of a triangle are equal to two right angles.  This fact, Lotze claims, is not subject to experimental verification or refutation.  If astronomical measurements of very large distances showed that the three angles of a triangle add up to less than two right angles, we would conclude that a hitherto unknown kind of refraction has deviated the light-rays that form the sides of the observed triangle.  In other words, we would conclude that physical reality in space behaves in a peculiar way, but not that space itself shows properties which contradict all our intuitions and are not backed by an exceptional intuition of its own.\footnote{``k\"{a}me es aber einmal dazu, da{\ss} astronomische Messungen gro{\ss}er Entfernungen nach Ausschlu{\ss} aller Beobachtungsfehler eine kleinere Winkelsumme des Dreiecks
nachwiesen, was dann? Dann w\"{u}rden wir nur glauben, eine neue sehr sonderbare Art
der Refraction entdeckt zu haben, welche die zur Bestimmung der Richtungen dienenden
Lichtstrahlen abgelenkt habe; d.h. wir w\"{u}rden auf ein besonderes Verhalten des
physischen Realen im Raume, aber gewi{\ss} nicht auf ein Verhalten des Raumes selbst
schlie{\ss}en, das allen unseren Anschauungen widerspr\"{a}che und durch keine eigene
exceptionelle Anschauung verb\"{u}rgt w\"{u}rde.''  \cite[pp. 248, 249]{Lotze} \cite[p. 408]{TorrettiPhilosophy} }  \cite[288, 289]{TorrettiPhilosophy} \end{quote}
The additional ingredients needed beyond Lotze were supplied by Neumann, Seeliger and Einstein's modification of Newtonian gravity.  Doing to General Relativity what those three had done to Newton's theory would restore a flat background space-time geometry, one difficult to observe due to gravity's almost-universal distortion effects, but observable on long distances due to the new term (eventually interpretable as a graviton mass) in the field equations.

%%%%%%%%%%%%%%%%%%%%%%%

\subsection{Neglect of Massive Gravity} 
  
Was massive gravity or something like it part of the discussion when philosophers were pondering General Relativity? It was not.  Probably the closest that one can find is a near-miss in the work of Peter Mittelstaedt \cite{MittelstaedtLorentz,Mittelstaedt}.  It is notorious in physics that general relativists and particle physicists do not tend to interact profitably regarding gravitational physics \cite[Preskill and Thorne foreword]{Feynman} \cite{Rovelli,BrinkDeserSupergravity}.  It is equally clear, if one knows what to look for, that the literature in the history and philosophy of space-time and gravity tends to ignore the particle physics side. Conveniently enough, one can identify a couple of sources that are sufficiently comprehensive that, if massive gravity had been part of the discussion, then those sources most likely would have noticed it.  Hentschel's massive study helpfully includes a section on other theories entertained at the time. One of the key features, at least to the particle physics-trained eye, is a privation:   nothing like massive gravity appears \cite[pp. 46-54]{Hentschel}.  Neither does the name of Markus Fierz, Pauli's collaborator in identifying the linearized source-free Einstein's equations as massless spin 2, appear in the index or in the bibliography. Another quite comprehensive source likely to mention massive gravity if it had arisen is Combridge's bibliography \cite{Combridge}, but evidently it did not arise in the literature of 1921-37.

If anyone had paid attention both to particle physics  and to the status of Kantian philosophy \emph{vis-a-vis} General Relativity, an obvious question would have been, ``why not do to Einstein's theory what Seeliger, Neumann and Einstein did to Newton's?''  Such a question should have been all the easier to ask once the exponentially decaying potential had a physical meaning as a graviton mass in the 1920s-30s, as opposed to the bare parameter in Neumann's, Seeliger's and Einstein's works.  One can see from Hentschel's work that no such thing happened \cite{HentschelKant,Hentschel}. Evidently even ideas in the neighborhood of Lotze's were not entertained much.  While the bibliography contain's Lotze's \emph{Grundz\"{u}ge der Naturphilosophie}, his name does even not appear in the exhaustive index, much less the relevant sections.
If we can excuse the historical actors of the 1920s-30s, it isn't necessary to follow them.

In one sense something like massive gravity was already part of the discussion, namely, Schlick's discussion of Seeliger \cite[ch. 9, p. 70]{Schlick}---but this section soon become scientifically obsolete.  Recognizing the problem of the divergent gravitational potential for an infinite homogeneous Newtonian universe and  Seeliger's solution to it, Schlick says only this:
``An unsatisfactory feature of this theory is, however, contained in the fact that the hypothesis
is invented ad hoc, and is not occasioned or supported by any other experience.'' 
Schlick, it is noteworthy, makes his own judgment of Seeliger rather than being misled by Einstein's false analogy.  Schlick's judgment is far from prophetic in insight, however.  While his complaint is true of many of Seeliger's \emph{ad hoc} force-laws, it is less true of the Neumann-Einstein modification of the Poisson equation.  More to the point, once the idea that matter is fields and hence must satisfy relativistic wave equations caught on---an idea with antecedents in the 1910s electromagnetic world picture and more clearly evident in Pascual Jordan in the late 1920s---the ubiquity of slow-moving matter (rocks, trees, tables, buildings, \emph{etc.}) implied that there was an enormous amount of experience supporting rest mass terms for matter fields at least by the late 1930s.  It is plausible by analogy that electricity or gravity might have the same feature.  Not coincidentally, such a development is one of the noteworthy features of 1920s-30s physics (de Broglie's massive photons, the Klein-Gordon equation, the Yukawa potential, Proca's massive vector meson field, Wigner's mass-spin taxonomy, \emph{etc}.).  Thus there an overwhelming amount of experience of matter described by relativistic wave equations with mass terms, and the articulated theoretical possibility due to de Broglie from 1922 that electromagnetism had the same feature.  Why shouldn't gravity also, a topic later considered in that light largely by Marie-Antoinette Tonnelat and Gerard Petiau on a sustained basis with de Broglie's involvement  \cite{TonnelatWaves,PetiauCR41b,PetiauCR41c,deBroglie1,TonnelatDisquisitiones,Tonnelat20,TonnelatGravitation,TonnelatNewtonian,PetiauRadium45,deBroglie}?

%%%%%%%%%%%%%%  

Prior to his underestimate of Seeliger, Schlick had already made his own job easier by adopting a policy of dismissing the type of underdetermination-by-approximation  worries that concerned Seeliger and Duhem.  The merely approximate character of the confirmation of Special Relativity was no obstacle to accepting the theory as exact for philosophical interpretation \cite[p. 159]{SchlickSignificance}! Such a claim seems to conflate interpreting a theory and interpreting the range of theories that fit the data at hand.

 Such a claim might make sense if one is convinced that there exist no plausible and conceptually different theories that approximate the theory in question.  The availability from the 1920s onward of the concept of massive theories, and especially recognition (probably from the 1930s \cite{FierzPauli,PauliFierz}) of their tendency to have smaller symmetry groups than do massless theories, made it appropriate to recognize what Schlick hadn't envisaged, namely, plausible and conceptually distinct theories that approximate the theory in question.  If the photon or graviton mass were just another really small parameter that might be $0$, it would be reasonable to ignore it.  But since the photon or graviton mass is a physically meaningful concept, indeed one of a type that is exemplified for at least some other fields (the electron field is massive, for example, as are the weak force bosons), and since a photon or graviton mass breaks the gauge symmetry `group' and hence makes a large conceptual difference,  while a scalar graviton mass at least breaks the conformal group and leaves only the Poincar\'{e} group, the conditions that license ignoring such rivals are not satisfied.  
 Schlick  \cite{Schlick,SchlickCritical}, though trained as a physicist and hence capable of making his own assessments, failed to see the potential significance of Seeliger's work. Yet analogous ideas to the Neumann-Seeliger mathematics would very shortly start to emerge independently in the work of de Broglie  \cite{deBroglieBlack,deBroglieWavesQuanta,deBrogliePhilMag} and others for massive particles/waves in the Klein-Gordon equation.  But Schlick's unrevised views remained influential in philosophy.

%%%%%%%%%%%%%%%%%%%%%%%%%%%%

\subsection{Schlick's Hymn to the Universe, General Relativity, and Einstein} 

By contrast,  Schlick's enthusiasm for what Einstein achieved with the cosmological constant $\Lambda$ \cite[pp. 70-75]{Schlick} is perhaps unparalleled by any subsequent writer, especially as shown in the hymn to the universe, General Relativity, and Einstein. 
\begin{quote} 
The structure of the universe, which the general
theory of relativity unveils to us, is astounding in its logical
consistency, imposing in its grandeur, and equally satisfying
for the physicist as for the philosopher. All the difficulties
which arose from Newton's theory are overcome; yet all the
advantages which the modern picture of the world presents,
and which elevate it above the view of the ancients, shine
with a clearer lustre than before. The world is not confined
by any boundaries, and is yet harmoniously complete in itself. It is saved from the danger of becoming desolate, for
no energy or matter can wander off to infinity, because 
space is not infinite. The infinite space of the cosmos has
certainly had to be rejected; but this does not signify such
sacrifice as to reduce the sublimity of the picture of the
world. For that which causes the idea of the infinite to inspire
sublime feelings is beyond doubt the idea of the endlessness
of space (actual infinity could not in any case be
imagined); and this absence of any barrier, which excited
Giordano Bruno to such ecstasy, is not infringed in any way.

By a combination of physical, mathematical, and philosophic thought genius has made it possible to answer, by
means of exact methods, questions concerning the universe 
which seemed doomed for ever to remain the objects of
vague speculation. Once again we recognize the power of
the theory of relativity in emancipating human thought,
which it endows with a freedom and a sense of power such as
has been scarcely attained through any other feat of science. \cite[pp. 74, 75]{Schlick} 
\end{quote}
 Schlick seems not to have been disposed to use his training as a physicist to cultivate  an unbiased range of scientific options for philosophical evaluation. 

His remarks  should be compared with another assessment of the cosmological constant $\Lambda$, namely, that it was difficult to interpret \cite{FMS,McCreaLambda,KerszbergInvented}.
The matter was well described by Freund, Maheshwari and Schonberg, who were not confused by Einstein's false analogy. 
\begin{quote}
In the ``Newtonian'' limit it leads to the potential equation,
$$\Delta V + \Lambda =  \kappa \rho.  \hspace{1.3in}  (1)$$
Correspondingly, the gravitational potential of a material point of mass $M$ will be given by
$$ V   = -\frac{1}{2} \Lambda r^2 - \frac{\kappa M}{r}. \hspace{1in}  (2) $$
A ``universal harmonic oscillator'' is, so to speak, superposed on the Newton law. The
origin of this extra ``oscillator'' term is, to say the least, very hard to understand.
\cite{FMS}  \end{quote}
These remarks are not a bit poetic, but they are entirely reasonable by the standards of the late 1960s, a much better informed time than the late 1910s.

%%%%%%%%%%%%%%%%

\subsection{Reichenbach and Carnap Did Not Alter the Situation} 

One might think of Reichenbach or Carnap  as taking Schlick's baton in paying ongoing philosophical attention to space, time and General Relativity.  But on the issues at hand, Carnap and Reichenbach help to explain the persistence of the problem. 
 Reichenbach, despite his impressive and sustained engagement with space-time theory and geometry throughout the 1920s, doesn't help much on this point.  He has great praise for Schlick as philosophically unbiased and displaying a sure understanding of physics \cite[pp. 36, 37]{ReichenbachPresentState}.  That is of course true if it means the absence of other philosophers' biases \emph{against General Relativity} and having a comparatively solid understanding of \emph{that theory}.  Schlick's effort to tie Einstein's theory to Machian relationalism \cite{Schlick} seems to leave little room for gravitational radiation, however.  % page number? 
 Reichenbach makes  no mention of Seeliger, Neumann, or Lotze.
  He remained intelligently engaged with space-time theory at least throughout the 1920s; his most serious work has appeared only in German and is hardly accessible \cite[appendix]{Reichenbach1929Zeitschrift,ReichenbachAngewandte,ReichenbachRaum}.\footnote{Recently an unpublished draft translation of the appendix of \emph{Philosophie der Raum-Zeit-Lehre}, omitted from the published translation \cite{ReichenbachSpace}, appeared online \cite{ReichenbachAppendix}.  Thanks are due to Marco Giovanelli and to the Hans Reichenbach Papers at the Archives of Scientific Philosophy in the University of Pittsburgh library.}  He turned his attention largely elsewhere during the 1930s \cite{SalmonReichenbachLogicalEmpiricist}).  Apparently he never noticed the relevance of particle physics and massive graviton theories.  The fact that many of Reichenbach's minor works could be collected under the title \emph{Defending Einstein} \cite{DefendingEinstein} reminds us of the climate of contention and Reichenbach's role therein.

Carnap thought that Reichenbach had said what needed saying and hence quit writing on space-time \cite[p. 957]{Carnap}. Carnap presumably was complicit in the secret elimination of the lengthy appendix for the English translation \cite{ReichenbachSpace}, depriving readers of early 1920s developments in metric-affine geometry (showing the affine connection  to be conceptually independent of the metric) and the question of chronogeometric significance. 
 Yet Carnap's preface suggests that nothing important had happened in the 30 years since the (longer) German original appeared.  With Carnap's \emph{imprimatur}, four decades of physics (roughly 1918-1958) were written off for philosophers of space-time.  The relevant science had terminated in Einstein's work, it now seemed.  

%%%%%%%%%%%%%%%%%%%%%%%%%%%%%%%%%%%%%%%%%%%%%%%%%%%%%%%%%%%

\section{On Friedman's Constitutive A Priori Role of the Principle of Equivalence}

If massive gravity renders unclear the need for Schlick's General Relativity-based anti-Kantian revolution, it also sheds light on Michael Friedman's recent claim that the principle of equivalence plays a constitutive \emph{a priori} role in General Relativity---that the principle of equivalence is required for the theory to have empirical content \cite{FriedmanDynamicsReason,FriedmanKantKuhn}.  (This critique of Friedman's treatment of the equivalence principle is complementary to Howard's critique \cite{HowardFriedman}, for we reach similar conclusions by different but compatible arguments.)   
I have in mind ``Einstein's principle of equivalence, which identifies gravitational effects with the inertial effects formerly associated
with Newton's laws of motion\ldots''  \cite[p. 37]{FriedmanDynamicsReason}.
\begin{quote} Such a variably curved space-time structure would have no empirical meaning or application, however, if we had not first singled out some empirically
given phenomena as counterparts of its fundamental geometrical notions --- here the notion of geodesic or straightest possible path. The principle of 
equivalence does precisely this, however, and without this principle the intricate
space-time geometry described by Einstein's field equations would not even be empirically false, but rather an empty mathematical formalism with
no empirical application at all.'' \cite[pp. 38, 39, footnote suppressed]{FriedmanDynamicsReason} \end{quote} 
Later he reiterates the point:
\begin{quote} in the absence of the principle of equivalence, Einstein's field equations
remain a purely mathematical description of a class of abstract (semi-)Riemannian
manifolds with no empirical meaning or application whatsoever. [footnote suppressed] \cite[p. 81]{FriedmanDynamicsReason}.  \end{quote} 
A bit later a weaker and more plausible claim is made:
\begin{quote} Einstein's field equations are thus logically possible
as soon as we have Riemannian manifolds available within pure mathematics,
but they are only really possible (possible as an actual description of some
empirical phenomena) when these abstract mathematical structures have
been successfully coordinated with some or another empirical reality. [footnote suppressed] \cite[p. 84]{FriedmanDynamicsReason} \end{quote}
While doubtless there is a job of coordination to do, and the principle of equivalence is a good way to do that job, the question is whether this principle of equivalence is unnecessarily strong.  In fact without the principle of equivalence as presented here, one could perfectly well test General Relativity if some weaker coordination principle were introduced, one that left gravity and inertia distinct.  One can compare  to massive spin-2 gravity, which one would expect to have nearly the same empirical content as General Relativity (for sufficiently small graviton mass, making the natural assumption of a smooth limit as the graviton mass goes to $0$) while differing radically from General Relativity on foundational issues \cite{FMS}.  Thus it is clear that the empirical content of General Relativity resides in the partial differential equations of the theory, not an additional principle about gravity and inertia.  At best the principle of equivalence (identifying gravity and inertia) might be a feature of the field equations of General Relativity, but it certainly does not needed to be added to the field equations.  

A \emph{prima facie} plausible philosophy of geometry for bimetric massive variants of (\emph{i.e.}, rivals to) General Relativity was outlined clearly by Freund, Maheshwari and Schonberg in the late 1960s in connection with their massive spin-$2$ gravitational theory \cite{FMS}.  The job of coordination gets done, but not by Friedman's principle of equivalence, which is clearly false for massive spin-$2$ gravity. Rather it is done by the field equations, gravitational and material, of the theory.  Such a theory, while strikingly different from Einstein's theory ontologically, approximates Einstein's theory arbitrarily well.  That is precisely analogous to what happens in de Broglie-Proca massive electromagnetism.\footnote{Actually various devils in the details arise for massive gravity, but one is hardly entitled to appeal to those until and unless one takes massive gravity seriously enough to see the point, and then  stares at it longer and harder to find devils in the details.   It is not as if those who didn't entertain massive gravity somehow ``knew all along'' that it didn't work. }  An illuminating but somewhat lengthy excerpt from that paper is included as an appendix.  Here is a small portion:
\begin{quote} 

\hspace{1.25in}  a) \emph{Breakdown of Geometrical Interpretation} \\ 

The theory, not being generally covariant, cannot be interpreted geometrically. This
means first of all that the quadratic form,
$$ d\sigma^2 = g_{\mu\nu} dx^{\mu} dx^{\nu} \,\,, $$
has nothing to do with the line element of the \emph{world} geometry, which remains
$$ds^2 = \eta_{\mu\nu}dx^{\mu} dx^{\nu} \,\, .$$

\ldots %Similarly, the equations of motion of matter\ldots still look formally as if they were geodesic equations. As a matter of fact, they are not. Indeed, the $\Gamma_{\mu\nu}^{\sigma}$ are given by the usual expressions, but $g_{\mu\nu}$ and\ldots  [its  inverse] are determined from the \emph{not}-generally-covariant equations\ldots. 
 The geometrical interpretation is one of the crucial steps in applications of Einstein's theory. What do we offer as a replacement? The field equations\ldots  and the equations of motion for matter\ldots fully
determine the answer to any question one can ask.\ldots % For that matter, this is true for Einstein's theory as well. There, however, geometrical considerations may be used as a luxurious shortcut toward the answers to many problems.

 \hspace{2in}   b) \emph{Local Problems} \\ 
If our theory is different from Einstein's, does this mean that it conflicts with the classical
tests of the latter? No. All classical tests are \emph{local}, i.e., they involve only small
regions of space and time. Locally our theory differs from that of Einstein only by
terms of the order (radius of system/Hubble radius),\ldots % so that the corrections are indeed negligible and the local tests cannot distinguish between the two theories. %Moreover,
%\emph{locally} one can reinstate an approximate geometrical interpretation.\ldots % One may wonder
%%whether there is any sense to an approximate gauge invariance. Fortunately, there is a
%%test case available in nature: chiral gauge invariance. Even though the breaking of the
%%gauge invariance occurs through a mass as large as that of the $\pi$-meson, the low-energy
%%theorems that follow from the chiral-gauge group are still valid to a very good degree of
%%accuracy. It is thus totally justified to expect the low-energy theorems of Einstein's
%%theory to hold to a much better accuracy as $\Lambda/m_{\pi}^2 = O(10^{-80})$. 
%Thus at a local level our theory is indistinguishable by usual experiments from that of Einstein. The real
%difference appears for systems of the size of $\Lambda^{-\frac{1}{2}},$ that is, for cosmological problems.
%%We shall discuss these in detail in the next section. Here let us only emphasize once more
%%that ours is a theory in flat space. The pseudo-Euclidean metric can be observed only in
%%cosmological experiments. Local experiments could detect it only if performed accurately
%%enough to be sensitive to terms of the order (size of system/Hubble radius).
 \cite{FMS}
\end{quote}
Thus the principle of equivalence is not necessary for empirical content even in Einstein's theory (at least if the principle of equivalence is something over and above Einstein's field equations and their coordination to gravity and heavy matter). Today's philosophical reader will sense some affinity with Brown's space-time philosophy \cite{BrownPhysicalRelativity}, especially because both attend to theories with more than one metric \cite{BrownFest}.

As it turns out, if one thinks carefully and consistently about causality in massive gravity, matters get complicated  \cite{MassiveGravity1}. 
 In fact many facets of massive spin-$2$ gravity get subtle on close enough inspection, problems that do not arise in electromagnetism.  But that is hardly a vindication of ignoring the theory and being (maybe) right for the wrong reason.  Sometimes in cartoons one can be  systematically lucky, as when unwittingly Bugs Bunny was chased by a hungry vampire in the latter's castle and happened to utter magic words at just the right times to avoid being bitten \cite{BugsVampire}. %(\emph{Transylvania 6-5000}, 1963).  
But space-time philosophy is not a topic in which one cannot rationally plan to be lucky.  One therefore needs to attend to alternative possibilities (unless one is Hegelian perhaps\footnote{On Hegel, Lakatos and Feyerabend, see (\cite{MotterliniLakatosArchive})}).

%%%%%%%%%%%%%%%%%%%%%%%%%%%%%%%%%%%%%%%%%%%%%%%%%%%%%%%%%%%%%%%%%%%%%%%%%%%%%%%%5

\section{`Massive' Newtonian Gravity Is Strictly Galilean}

There is an interesting irony for Kant's views on Newton's physics.  Kant argues, as described by Michael Friedman, that we 
\begin{quote} 
need to presuppose the immediacy and universality of gravitational attraction in order to develop a rigorous method for comparing the masses of the primary bodies in the solar system. [footnote suppressed]  We need such a method, in turn, in order rigorously to determine the center of mass of the solar system. \cite[p. 157]{FriedmanKantExact}
\end{quote}
That is important because
\begin{quote} \ldots Kant does not have the concept of inertial frame and instead views
the Newtonian laws of motion (together with other fundamental principles Kant takes to be a
priori) as defining a convergent sequence of ever better approximations to a single privileged frame of reference (a counterpart of absolute space) at rest at the center of gravity of all matter. \cite[p. 37]{FriedmanDynamicsReason} \end{quote} 
 Kant also appears to say (though on balance Friedman thinks otherwise \cite[pp. 166, 167]{FriedmanKantExact}) that the  $\frac{1}{r^2}$ law is \emph{a priori} due to geometry.  

Kant's view is, in light of 20th century particle physics, almost backwards.   A  $\frac{1}{r^2}$ force comes from Laplace's equation in spherical symmetry in three spatial dimensions.  But since Neumann's work it has become clear that there is a more general way to have a 3-dimensional equation akin to Laplace's, but with a new parameter---what one would now call a graviton mass.  Laplace's equation is only appropriate for massless gravitons.  If space is three-dimensional but the graviton has a small mass $m$, then gravity has instead a $\frac{e^{-mr} }{r}$ potential instead of   $\frac{1}{r}$.  The force is again given by the derivative of the potential.  Thus the geometrical argument from the 3-dimensionality of space and solving a Laplace-like linear differential equation excludes many possible force laws (including some of Seeliger's), but does not count against $\frac{e^{-mr} }{r}$.  Hence  a $\frac{1}{r}$ potential cannot be known \emph{a priori}.

Things get a bit worse for Kant's views of what Newtonian gravity allows one to know.  The symmetry group of Newtonian gravity is larger than the Galilean group and contains accelerations \cite[p. 294]{MTW} \cite{SmithNewton} \cite[p. 424]{Newton}.  This isn't really news, given that Newton said as much (although Kant apparently struggled even with the  Galilean relativity symmetry, to say nothing of the less famous symmetries larger than the Galilean group).  Newton wrote:
\begin{quote} 
Corollary 6. If bodies are moving in any way whatsoever with respect to one another and are urged by equal accelerative forces along parallel lines, they will all continue to move with respect to one another in the same way as they would if they were not acted on by those forces.  \cite{SmithNewton} \cite[p. 424]{Newton} 
\end{quote}  
Hence one cannot tell using observations of the solar system whether the whole system is accelerating or not, much less whether it is at rest, \emph{pace Kant}.  

But massive graviton theories bring Kant some more good news:  one can get more of what Kant wanted from Newton's theory if one uses Seeliger-Neumann-Einstein ``massive Newtonian gravity'' (if the reader will permit the anachronism, which comes naturally to particle physicists \cite{DeserMass}).  The graviton mass term, which is algebraic in the gravitational potential, destroys the symmetries beyond the Galilean group.  Kant's lacking the concept of an inertial frame of reference and believing in a preferred frame in which the center of mass is at rest \cite[p. 37]{FriedmanDynamicsReason} %  
leave him destined for disappointment by any theory with a boost symmetry, whether Galilean or relativistic.  But by having only the Galilean symmetry group, massive Newtonian gravity comes much closer to achieving Kant's goals than does Newton's theory.  That Newton's theory isn't a necessary truth turns out to be perhaps a good thing for Kant.  

% Fries Bern Buldt? 

%%%%%%%%%%%%%%%%%%%%%%%%%%%%%%%%%%%%%%%%%%%%%%%%%%%%%%%

\section{Conclusion}

One cannot rightly understand the actual philosophical significance of General Relativity, including the true rational force of  its destructive impact on neo-Kantianism and the rationality of the views of the leader of the Vienna Circle, without attention to particle physics.  If anything \emph{really} made a neo-Kantian philosophy of geometry impossible (to a scholar who transcended the usual disciplinary boundaries but required no superhuman intelligence), it happened in the 1970s \cite{vDVmass1,vDVmass2,DeserMass}, when massive spin-$2$ gravity died (or at any rate seemed to die).   
 Finally one had to accept the conceptual innovations of General Relativity, half a century after Schlick had claimed so on much weaker grounds. This overthrow of Kant was entirely unheralded at the time.  Few philosophers still cared by the early 1970s about a Kantian philosophy of geometry.  Philosophers sought no guidance from particle physicists about space-time (despite its potential value by then \cite{FMS}).  Particle physicists paid little attention to philosophy  \cite{MerminFeynmanShutUpandCalculate}. % . 
  But if the philosophy of geometry is not to be  held captive by historical accidents, then the cause of death for a Kantian philosophy of geometry---in rationally reconstructed history!---involved the van Dam-Veltman-Zakharov discontinuity of massive pure spin-$2$ gravity in the limit of $0$ graviton mass and the threat of instability (but see (\cite{MaheshwariIdentity,MassiveGravity3}) on the latter point). %
 This philosophical death is also apparently reversible, and perhaps now reversed \cite{Vainshtein2,deRhamGabadadze,HassanRosen,HinterbichlerRMP,deRhamLRR}. Massive spin-$2$ gravity might live, at least for now. (So it has seemed to a fair number of working physicists within our own decade.)  So might synthetic \emph{a priori} knowledge live, if one wants it to. I do not write to defend it, but to show that physics has left the matter open until more recently than is generally believed, and that if and when physics forecloses the option, the grounds will be different from Schlick's.  
  Of course the attention focussed on massive spin-$2$ gravity might wind up hastening its demise \cite{DeserWaldronAcausality}; if fatal objections are there to be found, they will be found faster now that people are looking.  Maybe Kant's synthetic \emph{a priori} knowledge is finally being scientifically refuted definitively a century after General Relativity appeared?  Perhaps a good argument will vindicate Schlick's claims at last.  

Particle physics has also proven useful recently in the historiography of General Relativity, shedding light on Einstein's invocation of energy conservation and on what was really wrong with his 1913 \emph{Entwurf} theory \cite{EinsteinEnergyStability}. 

%%%%%%%%%%%%%%%%%%%%%%%%%%%%%%%%%%%%%%%%%%%%%%

\section{Appendix:  Philosophy of Geometry from Massive Spin 2 Gravity}

Freund, Maheshwari and Schonberg comment on their theory as follows: 
\begin{quote} 
It is the [non-generally covariant or bimetric part of the graviton mass] term that contains all the novel features of our theory. Without it the Lagrangian
would lead to generally covariant field equations and as such would describe a massless
field [\emph{i.e.}, General Relativity]. It is only the presence of this one term that breaks general covariance. The departures
from Einstein's theory can now easily be identified.

\hspace{1.25in}  a) \emph{Breakdown of Geometrical Interpretation} \\ 
\vspace{-.11in}

The theory, not being generally covariant, cannot be interpreted geometrically. This
means first of all that the quadratic form,
$$ d\sigma^2 = g_{\mu\nu} dx^{\mu} dx^{\nu} \,\,, $$
has nothing to do with the line element of the \emph{world} geometry, which remains
$$ds^2 = \eta_{\mu\nu}dx^{\mu} dx^{\nu} \,\, .$$

Similarly, the equations of motion of matter\ldots  still look formally as if they were geodesic % suppressed equation numbering ``(25)''
equations. As a matter of fact, they are not. Indeed, the $\Gamma_{\mu\nu}^{\sigma}$ are given by the usual
expressions, but $g_{\mu\nu}$ and\ldots  [its  inverse] are determined from the \emph{not}-generally-covariant [field] equations\ldots, so that the $\Gamma_{\mu\nu}^{\sigma}$ are not genuine Christoffel symbols. The geometrical interpretation is one of the crucial steps in applications of Einstein's theory. What do we offer as a replacement? The field equations\ldots and the equations of motion for matter\ldots fully
determine the answer to any question one can ask. For that matter, this is true for
Einstein's theory as well. There, however, geometrical considerations may be used as a
luxurious shortcut toward the answers to many problems.

 \hspace{2in}   b) \emph{Local Problems} \\ 
If our theory is different from Einstein's, does this mean that it conflicts with the classical
tests of the latter? No. All classical tests are \emph{local}, i.e., they involve only small
regions of space and time. Locally our theory differs from that of Einstein only by
terms of the order (radius of system/Hubble radius), so that the corrections are indeed
negligible and the local tests cannot distinguish between the two theories. Moreover,
\emph{locally} one can reinstate an approximate geometrical interpretation.  One may wonder
whether there is any sense to an approximate gauge invariance. Fortunately, there is a
test case available in nature: chiral gauge invariance. Even though the breaking of the
gauge invariance occurs through a mass as large as that of the $\pi$-meson, the low-energy
theorems that follow from the chiral-gauge group are still valid to a very good degree of
accuracy. It is thus totally justified to expect the low-energy theorems of Einstein's
theory to hold to a much better accuracy as $\Lambda/m_{\pi}^2 = O(10^{-80})$. 
Thus at a local level our theory is indistinguishable by usual experiments from that of Einstein. The real
difference appears for systems of the size of $\Lambda^{-\frac{1}{2}},$ that is, for cosmological problems.
%We shall discuss these in detail in the next section. 
\ldots Here let us only emphasize once more that ours is a theory in flat space. The pseudo-Euclidean metric can be observed only in
cosmological experiments. Local experiments could detect it only if performed accurately
enough to be sensitive to terms of the order (size of system/Hubble radius).
 \cite{FMS}
\end{quote}

As noted above, there arose devils in the details in the early 1970s, which might or might not have been exorcised recently.  Exactly this relationship holds, however, in the simpler scalar case between massless spin $0$ (Nordstr\"{o}m's 1914 theory) and massive scalar gravity \cite{DeserMass,PittsScalar,ScalarGravityPhil}, as Seeliger already expected in the 19th century.  
% published version says "in the already 19th century." :-(

%%%%%%%%%%%%%%%%%%%%%%%%%%%%%%

\section{Acknowledgments}

I thank Marco Giovanelli, Dennis Lehmkuhl, and Jeremy Butterfield for assistance and discussion, Karl-Heinz Schlote for helping me to find some of Neumann's works,  Don  Howard for much useful instruction, and the anonymous referees for helpful comments. This work was supported by John Templeton Foundation grant \#38761.

% 

%\bibliographystyle{apalike}
%%
%\bibliography{Pitts} 
%\end{document}

\end{document}